# ScaleSimulator – A Fast and Cycle-Accurate Parallel Simulator for Architectural Exploration

Ori Chalak
Huawei,
Israel
ori.chalak@huawei.com

Cai Weiguang
Huawei,
China
caiweiguang@huawei.com

Li Wei
Huawei,
China
jacky.liwei@huawei.com

Fang Lei
Huawei,
China
fanglei3@huawei.com

Zheng Libing
Huawei,
China
zhenglibing@huawei.com

Wang Jintang
Huawei,
Israel
wangjintang@huawei.com

Wu Zuguang
Huawei,
China
wuzuguang@huawei.com

Gu Xiongli
Huawei,
China
guxiongli@huawei.com

Wang Haibin
Huawei,
China
benjamin.wanghaibin@huawei.com

Avi Mendelson
Technion
Israel
avi.mendelson@technion.ac.il

## ABSTRACT

Design of next generation computer systems should be supported by simulation infrastructure that must achieve a few contradictory goals such as fast execution time, high accuracy, and enough flexibility to allow comparison between large numbers of possible design points. Most existing architecture level simulators are designed to be flexible and to execute the code in parallel for greater efficiency, but at the cost of scarified accuracy.

This paper presents the ScaleSimulator simulation environment, which is based on a new design methodology whose goal is to achieve near cycle accuracy while still being flexible enough to simulate many different future system architectures and efficient enough to run meaningful workloads. We achieve these goals by making the parallelism a first-class citizen in our methodology. Thus, this paper focuses mainly on the ScaleSimulator design points that enable better parallel execution while maintaining the scalability and cycle accuracy of a simulated architecture.

The paper indicates that the new proposed ScaleSimulator tool can (1) efficiently parallelize the execution of a cycle-accurate architecture simulator, (2) efficiently simulate complex architectures (e.g., out-of-order CPU pipeline, cache coherency protocol, and network) and massive parallel systems, and (3) use meaningful workloads, such as full simulation of OLTP benchmarks, to examine future architectural choices.

## 1 INTRODUCTION

Simulators are essential tools for designing future systems and exploring architectural tradeoffs; they are used to evaluate new features, search for optimal configurations of hardware before it is manufactured, develop software layers of non-existing systems, and more. A major challenge in developing such simulation environments is the need to use current generation hardware to evaluate future generations of systems, which are most likely be significantly larger, faster and with advanced functionality. On top of all these “system obstacles,” such simulation environments are expected to be able to execute legacy software as well as future software layers that most likely require new and different optimization points. To that end, simulators must be able to trade between accuracy, execution time and flexibility. Traditionally, it was assumed that only two of these three goals can be achieved by the same tool at the same time [1]. This paper presents a new simulator, called *ScaleSimulator*, and a software environment that aims to achieve a different design point: the ScaleSimulator was developed to support the development of commercial products. Hence, it was required to provide a full system simulation, be cycle accurate while still being flexible, and be fast enough to run meaningful benchmarks, such as OLTP.

The ScaleSimulator is a modular simulation environment, easy to build and modify. It allows the entire system to be modeled “almost” at the micro-architectural level, while still running fast enough to simulate a massive multiprocessor system consisting of tens of cores, running meaningful workloads such as Hadoop [2], OLTP [3] or SPEC2006 [4]. At the heart of the proposed new methodology and tool is the new parallelization technology we developed. Most of this paper is thus devoted to this crucial technology, which makes it possible to achieve fast and efficient parallel execution time while still maintaining accuracy and flexibility. We will discuss how to achieve efficient execution time when simulating a massive number of processors, even at the data center level, and how to expedite the execution time while simulating a relatively small number of complex and detailed single-core architectures. Please note that although the work presented here is discussed in the context of the simulation environment, we believe it is applicable to other parallel systems as well.

### 1.1 Background and related works

Traditionally, a computer system is simulated either by using cycle accurate simulators [5] [6] [7] [8] [9] or by using event based simulation [10] [11]. Whereas cycle accurate simulation can achieve maximum accuracy at the cost of slower execution time due to the poor parallelization of the code, parallel discrete event simulation (PDES) exhibits less than the expected scalability due to frequent synchronization points [12] [13] [14] [8]. To achieve better scalability, some works suggest using relaxed synchronization. This approach could significantly increase the actual parallelism of the simulator, but may not be able to accurately and efficiently simulate complex architectures such as out-of-order cores and cache coherency protocols, where events may need to maintain a specific order of execution and response. Zsim [15] uses a two-phase approach termed "bound and weave"; during the bound phase, each core can be simulated w/o taking into account the execution of other cores. Thus, to accurately simulate many cores that share resources, each core records all events that may be of interest. During the second phase, the weave cycle, the simulator executes all events that were recorded during the first phase, and adjusts the execution time of all the cores. This approach allows hundreds or even thousands of cores to be simulated. The authors of the paper claim that for the benchmarks they tested, the accuracy of the model was very high.

Our design points were quite different. Since the goal of the simulation is to assist the architects of future products, we were asked, on the one hand, to be cycle accurate, and on the other hand to be as scalable as if we were using a relaxed synchronization model, so we could run meaningful benchmarks on realistic hardware configurations.

As we will not be able to cover all the different aspects of our simulation environment, this paper focuses mainly on how we achieve these two goals. We will begin with a high-level description of our simulation environment, followed by a detailed description and discussion of our unique approach to maintaining scalability and accuracy. Next, we will show a few of our simulation results that we believe are indicative of the overall scalability of our new proposed simulation tool. We end the paper with conclusions and directions for future work.

## 2 The general structure of our simulator

This section provides a high-level description of and introduction to the general structure of our proposed simulation environment. Please note that since our main focus is the scheduling methodology and parallel execution, this section will only give a general overview that we hope will be sufficient for understanding the challenges we faced when trying to parallelize the execution.

At the highest level, the simulation environment is divided into two parts (as indicated by Figure 1); the functional model (FM) and the performance model (PM) (similar to [13] [15]). The functional model represents a correct possible execution of the multicore environment. It may or may not run a "full system", that is, include the impact of the operating system. Please note that when running a parallel program on a multi-core environment, a different execution path might be obtained for different runs. So, the FM is required to generate a legal execution path of each core, and if possible to ensure that this path can represent the average case. In CPU simulation the functional model may use an emulator such as QEMU [16] as its base, since is proven to be extremely fast, supports cross platforms, and runs unmodified OS and software stack. Such an emulator might be used, e.g., to simulate an ARM-based system, while the host platform runs on x86 based CPUs. Please note that the functional part is almost independent of the performance model and so can easily be replaced by other tools; e.g., when appropriate, we use synthetic workloads.

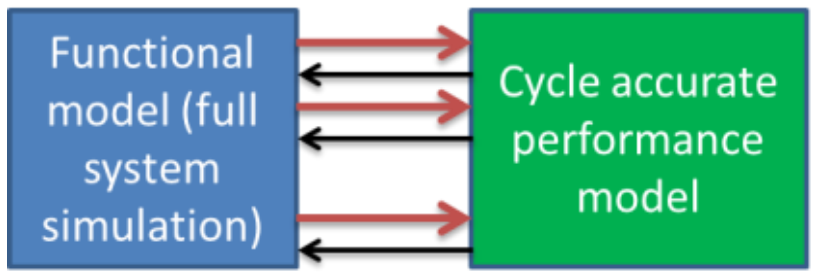

**Figure 1: General structure of the simulator**

The performance model we use here is somewhat similar to SystemC TLM 2.x [17] [9]. The basic entities of the system are clock, units, ports and messages (see Figure 2), where a unit stores its state and implements the timing aspect of the model and ports to communicate messages between units.

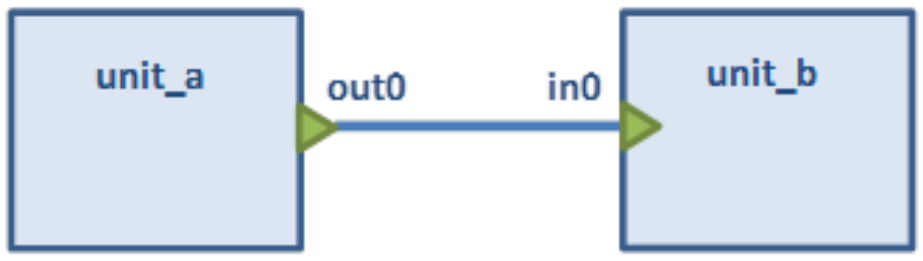

**Figure 2: Basic structures of the performance model**

The operation a unit needs to perform may last several cycles; it can be pipelined and may contain internal resources as well. The operation of a unit is driven by messages arrived to input ports. It submits the result messages to output ports and may be blocked if its output ports are blocked. A port represents a connection between units; it may also contain meta-data such as capacity, delay, etc.
Since the paper doesn't focus on how the model is constructed, we assume reader familiarity with these basic notations, or suggest getting them from many related sources (such as [6] [17]).

## 3 Parallel execution of the *ScaleSimulator*

To achieve a fast and cycle-accurate system level simulator, *ScaleSimulator* aims to take maximum advantage of the massive parallelism, many-core architecture provided by, e.g., [18] [19]. The method we use in this work is based on the notion of "design for parallelism", meaning that we specify the model, building all data-structures and using algorithms that ensure efficient execution of the system on parallel architectures. Our methodology, which we term 2.5-Phase Design, guarantees a thread-safe lockless data access. That is, we assume that all operations executed by the simulated system within a cycle are emulated using the following two and a half phases:

- Work phase: all threads perform in parallel the unit's computation of this cycle
- A short synchronization phase implemented using a software barrier
- Transfer phase: all threads perform in parallel the message transfer between units

- A short synchronization phase implemented using a software barrier

In general, this model preserves the "cycle accurate nature" of our simulator. An exception is a naïve use of the model that may force us to compromise on accuracy and to diverge from the actual hardware implementation, as happens when operations such as reads and writes to register files are performed in the same cycle. We examined the impact of this event on the overall performance and found it to be less than 1%. But if accuracy of the hardware model is required, then a possible work-around is to multiply the clock. Such a method will solve the problem, but is beyond the scope of this paper.

Using this methodology helps us to achieve the main goal of our tool – provide a flexible and easy-to-construct cycle-accurate simulation environment capable of simulating a massive number of future parallel cores while still being fast enough to simulate a meaningful set of massively parallel workloads.

### 3.1 Enabling parallelism through the design rules of a simulated model

To allow the amount of parallelism to be maximized, we built the system's models so that:

(1) Each hardware model is implemented as a unit. The system aggregates a cluster of units into a single thread.
(2) All operations are assumed to be executed in one cycle, meaning that if an operation takes, for example, 3 cycles, we will simulate it as 1-cycle operation followed by 2-cycle delay.
(3) A message sent in cycle $m$ will be consumed in cycle $n$ where $n > m$.
(4) Control and data are transmitted from one unit to another by messages.
(5) The messages are "transferred" over ports.
(6) All ports are connected point-to-point, making the data transfer contention free.

These rules enable a thread-safe parallelism of arbitrary model granularity, with completely lockless data access and, under all simulation scenarios, including back pressure, as if it is simulated in a serial manner (see Section 3.3).

### 3.2 Scheduling and Phases of execution

The proposed 2.5-phase execution calls to implement the system with two execution phases, using massively parallel execution resources and two (short) synchronization phases.

As we need to support a cycle-accurate simulation model, the ScaleSimulator works in a lock-step manner, executing the following phases for each simulated clock cycle.

#### 3.2.1 The "work" execution phase

During the work execution phase, all units that are ready to perform an operation will execute it. Please note that all operations being executed at the same "work cycle" are guaranteed by design to be independent of one another and so can be executed in parallel; the order of their execution does not impact the simulated results.
The work phase is typically composed of the following steps:

- Read input messages
- Read stored data
- Check output port vacancy
- Compute results
- Store results
- Submit results to output ports

For example, for a simplified CPU core, the unit that models the dispatch pipeline stage will do the following:

- Read (a) new instructions received from the front-end pipeline; (b) back pressure messages from the execution units.
- Store newly arrived instructions.
- Read stored pending instructions.
    - Check the vacancy of output ports driving the execution unit.
    - Decide which instructions to submit to which execution unit.
    - Submit the selected instructions to the selected output ports.

The use of point-to-point connections together with the design rules described before guarantees that no contention on resources occurs during that execution phase.

#### 3.2.2 The "transfer" execution phase

To guarantee the locality of references to resources, each operation writes its results to local buffers (done during the "work" phase), and the transfer phase is used to copy the pointer to the message from the output port to the receiver input ports. When the copy operation is complete, the input port is considered ready.

All operations executed during this phase can be fully parallelized among units. Please also note that we are moving pointers and not the message itself. This reduces significantly the amount of data copy and in turn contributes to the simulation speed.

We believe that the implementation presented in this paper could be further improved, e.g., by taking into consideration the hardware locality, similar to [7]. We are leaving the discussion on the potential of this optimization to future works.

#### 3.2.3 The barrier execution phase

The "half" phase of the execution aims to synchronize the system and ensure that the simulation progresses step-by-step to achieve a cycle-accurate behavior. The implementation of this section is crucial to the overall system performance, and thus will be discussed in depth in Section 4.

### 3.3 Back pressure

Before our model can take full advantage of the parallel architecture, the crucial issue of back pressure in an architectural-level simulator must be addressed. Back pressure is a scenario where part of the modeled architecture is blocked for further execution because a receiver unit slowed down or stopped and cannot receive more messages. For example, back pressure may occur in a simulated pipeline architecture when an execution unit is stalled due to read-after-write register dependency. This in turn may result in all the input queues of this unit being full, thus preventing the execution of an operation at an earlier stage of the pipeline. SimpleScalar [1] [20] handles back pressure by traversing the pipeline from the last stage to the first, so the back-pressure information can be propagated along this direction. This method suffers from sensitivity to the order of execution on the one hand and precludes efficient parallelism on the other.

ScaleSimulator strives to allow maximum parallelism and support any arbitrary execution order of units. The simulation result, either with respect to timing or with respect to computation, is indeed agnostic to the order of execution. Thus, we take a different approach and follow the rule: all back-pressure conditions of clock N must be calculated at cycle N-1 or earlier.

We do so by defining a set of "back pressure" ports that can be triggered at cycle N-1 in order to indicate that the system will have to be stalled at cycle N, as shown in Figure 3. By doing that, we can prevent the detection and the operation from being performed in the same cycle.

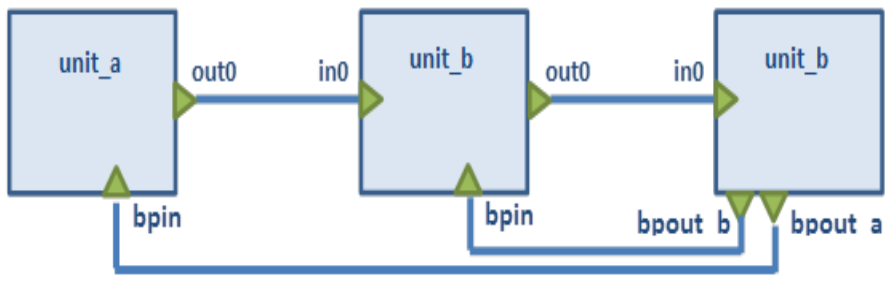

**Figure 3: Explicit back pressure**

Hence we can define two different ways to create a backpressure scenario: (1) explicit – as indicated above, dedicated ports that pass back-pressure messages to the units needing to be stalled (2) implicit – as any transfer occurs only if the receiver input port is vacant, an occupied input port causes the transfer to fail and the message remains in the sender's output port. In turn, the sender's output port remains occupied at the next cycle and prevents the sender unit from submitting more messages, which in turn causes the sender unit to stall. In the implicit back-pressure method, the back pressure ripples backwards cycle by cycle.

## 4 New approach for synchronization

Synchronization is one of the main limiters of the actual parallel scaling that can be achieved from parallel systems in general and from parallel simulators in particular. Event driven simulators are known to achieve better effective performance since they can "skip" simulating idle cycles, but the mechanism itself may not scale well in parallel. Thus, systems that need to "synchronize" too often (e.g., Hornet [9]) to achieve high accuracy present very poor scalability. Most modern simulation tools (e.g., [12] [15]) are therefore based on "relaxed synchronization", which introduces inaccuracy to the simulation but could speed up its execution time.

Our design point is different; the ScaleSimulator aims to build a cycle-accurate simulation of future architectures that may include a large number of cores. It should run efficiently enough to be able to simulate the run of meaningful workloads, such as SQL OLTP, for long enough to allow potential performance bottlenecks to be identified. It should also serve to assist in the choice of run sets of features that the company may implement on its future platform.

We observed that a major reason for using semaphores and other synchronization mechanisms is to assist in the "data-flow" based execution of operations, i.e., protect input and output ports or shared resources, such as common data structures. We propose, in contrast, a different model for the use of synchronization primitives:

- We use only software barriers to force different execution phases to be executed synchronously.
- We assume that all operations performed within the execution phase are race-free and need not be protected.

As we will show later, this new notion of synchronization allows us to achieve much better parallelism and to efficiently parallelize the execution of the cycle-accurate simulator.

Our model relies on the following two mechanisms: (1) a two-level scheduler and (2) a new barrier based mechanism that we termed a ladder-barrier.

Our goal is to simulate N units, called simulated units (SU) running on M physical cores (PC). In this paper we will assume that N=x*M where x≥1. Furthermore, our design methodology does not require that the user fully understand the details of the underlying implementation. In particular, the user will not have to change data structures or code in the simulator when changing the simulated environment.

In order to allow flexibility, we assume that the same simulation environment is being used regardless the number of simulated units or the number of physical cores selected for a specific run.

To this end, we allow the system to group the units into (M-1) clusters, where each group runs on a different physical core and is managed by a different internal scheduler, while the M'th core is dedicated to managing the simulation resources, services, and in particular, the higher-level scheduling and synchronization of the clusters.

At that point, we assume that the local scheduler, which runs all the units in its cluster sequentially, is in charge of running all events (work and transfer related) that belong to all simulated units assigned to the same physical core. The local scheduler is also in charge of communicating with the global scheduler, which runs on core M.

Figure 4 presents the two-level scheduling work flow, using the global and local schedulers. The algorithm of the global scheduler is described in the next section under the "ladder-barrier" synchronization mechanism.

```
while (true)
  for each cluster do in parallel
    work phase:
    for each unit in cluster do in serial
      unit.work()
    end
    barrier

    transfer phase:
    for each unit in cluster do in serial
      unit.transfer()
    end
    barrier
  end
end
```

**Figure 4: Two level scheduling**

As an example, consider the simple performance model in Figure 5.

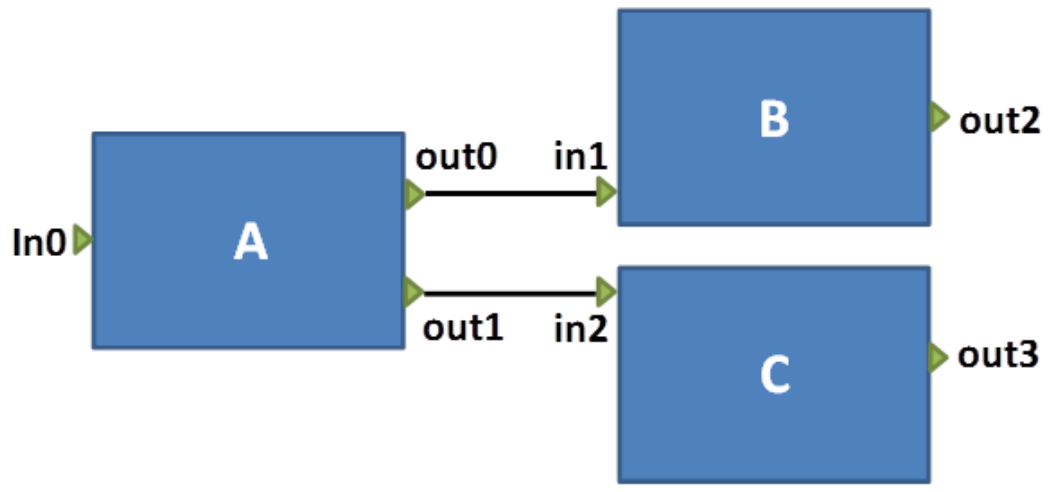

**Figure 5: A simple model**

The simulation of this model is parallelized among 3 worker threads, each assigned to simulate a cluster of one unit:

| **Thread** | 0 | 1 | 2 |
|---|---|---|---|
| **units** | A | B | C |

**Table 1: Simple model cluster assignment**

We use software barriers for time division of cycles and phases. Thread safety is guaranteed by time division. Thus, during each phase, each data item has a single owner that can safely read and write it, as can be seen in Table 2:

| **Thread** | 0 | 1 | 2 |
|---|---|---|---|
| **Work phase** | A,**in0**,out0 out1 | B,**in1**,out2 | C,**in2**,out3 |
| **Transfer phase** | out0,out1 **in1**,**in2** | out2 | out3 |

**Table 2: Simple model data ownership**

As can be seen, the input ports and message content switch the thread ownership between work and transfer and between transfer and work. For example:

- During the work phase, the message content is filled by thread 0 and the message pointer is written by thread 0 to out0.
- During the next transfer phase, this pointer is copied by thread 0 from out0 to in1.
- During the next work phase, thread 1 reads this pointer from in1 and the message content.

### 4.1 The ladder-barrier synchronization mechanism.

The ladder scheduler aims to define an efficient and scalable algorithm that can be used to efficiently schedule any number of clusters (physical cores) to run in a phase lock manner. In this work we assume that the global scheduler is running on a separate and dedicated core. We dedicate a separate core for the scheduler thread. The scheduler thread is in a wait state while the worker threads work, and can use this time for short maintenance tasks.
Sync-point is a primitive variable that enables an exclusive access by multiple threads. Practical examples are mutex, futex, semaphore, spinlock and atomic variable.

**Sync-point variables and operations:**

Each sync-point is common to two threads, where only one of them is a writer:

- Scheduler thread
- One of the worker threads

Operations of a specific sync-point of a specific thread:

- lock(sync-point, thread)
- unlock(sync-point, thread)
- wait(sync-point, thread)

A common sync-point may be implemented using a multiplicity of sync-points between scheduler and individual worker, or it may be implemented for all worker threads where the scheduler thread is an exclusive writer (as in the common-atomic method below):

- lockAll(sync-point)
- unlockAll(sync-point)
- waitAll(sync-point)

For each sync-point, there is an exclusive writer thread:

| **sync-point** | **(un)locked by** | **waited by** | **barrier before** |
|---|---|---|---|
| WORK | scheduler | worker | work phase |
| TRANSFER | scheduler | worker | transfer phase |
| PHASE0 | worker | scheduler | phase 0 |
| PHASE1 | worker | scheduler | phase 1 |

**Table 3: The role of each sync-point**

The algorithm for the scheduler thread is given in Figure 6 and the algorithm for the worker thread in Figure 7.

```
tick()  begin
   lockAll(TRANSFER)
   unlockAll(WORK)
   waitAll(PHASE0)

   lockAll(WORK)
   unlockAll(TRANSFER)
   waitAll(PHASE1)
end

run(numCycles)  begin
   lockAll(WORK)
   lockAll(PHASE0)
   for each thread do
      invoke OS-thread
      call task(thread)
      thread.stop = false
   end
   for cycle from 1 to numCycles  do
      tick()
   end
   for each thread,  thread.stop = true
   unlockAll(WORK)
end
```

**Figure 6: Scheduler thread algorithm**

```
task(thread)  begin
   wait(WORK, thread)
   unlock(PHASE1, thread)
   while not thread.stop do
            work(thread)   // do real work
            lock(PHASE1, thread)
            unlock(PHASE0, thread)
            wait(TRANSFER, thread)

            transfer(thread)   // do transfer
            lock(PHASE0, thread)
            unlock(PHASE1, thread)
            wait(WORK, thread)
   end
   unlock(PHASE0, thread)
end
```

**Figure 7: Worker thread algorithm**

The outcome flow for ladder-barrier synchronization is demonstrated in Figure 8.

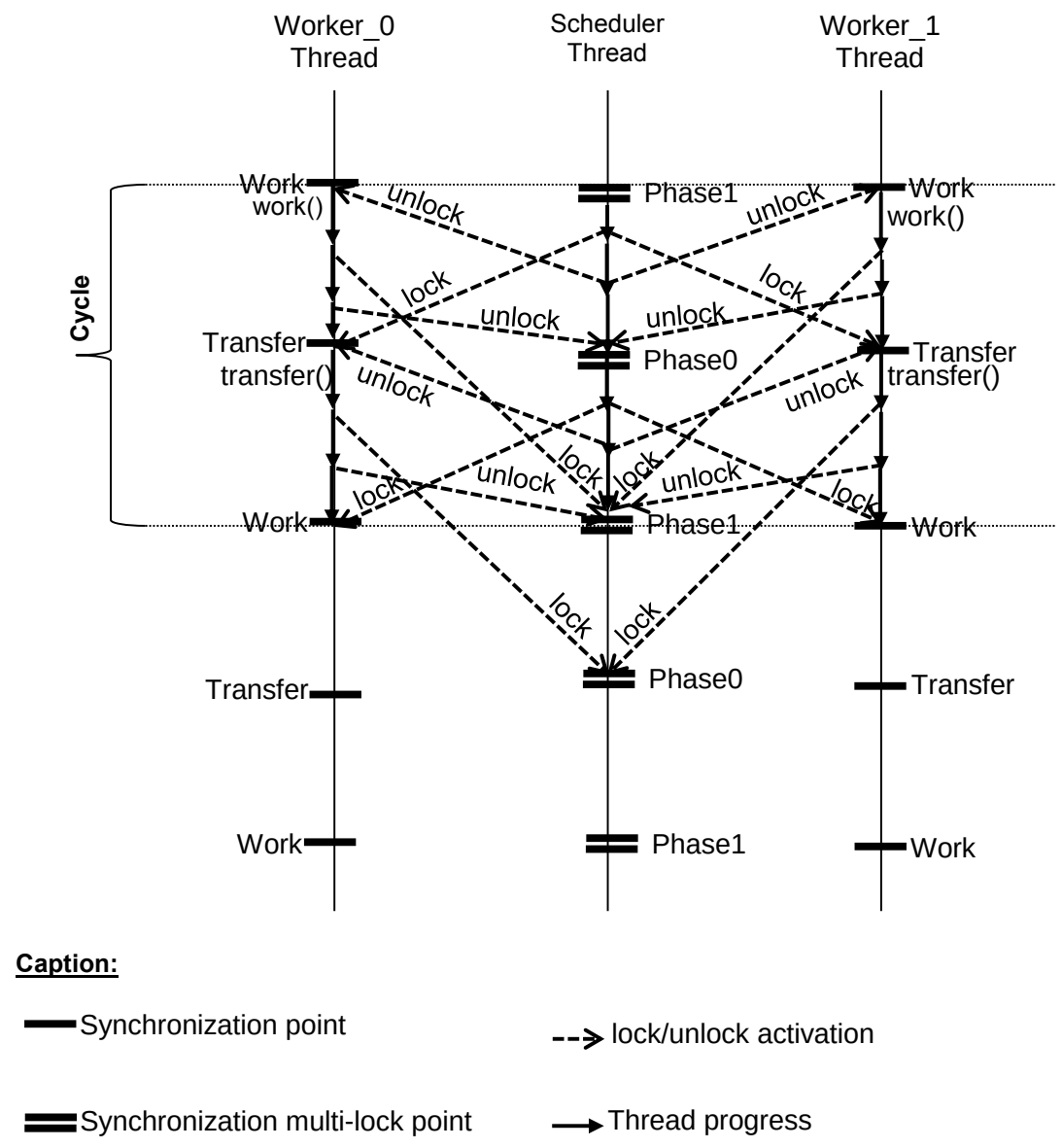

**Figure 8: Ladder-barrier synchronization**

As one can see, this mechanism can force all clusters to work in clock step and thus very efficiently implement the barrier synchronization between all clusters.

Please note that the worker needs no information on other clusters running on the system at any given time. Moreover, it is sufficient for the GS to have a table of all clusters that need a service at a given time. This table may be changed cycle by cycle to enable further optimization.

In the next section we will examine different ways to implement the sync-points and discuss different optimizations.

## 5 Evaluation results

We will start this section with a detailed discussion and evaluation of the new proposed synchronization mechanisms since they provide an upper bound on the overall scalability we can expect from the system. Next, we will evaluate our system using a simple in-order core running OLTP applications and when simulating a light data center model. We will conclude this section with simulation r Our goal is to simulate N units, called simulated units (SU) running on M esults of out-of-order cores running OLTP and spec based applications.

Please note that, due to the variability of the run-time results when using parallel systems, we run each experiment a few times and eliminate the extreme results. Also note that some experiments use different hardware configurations. We will indicate the relevant parameters for each experiment.

### 5.1 Synchronization methods

An important factor in measuring the efficiency of our implementation is the overhead incurred by synchronization primitives in general and barrier implementation in particular. We therefore implement the different primitives described in Figure 8 using 4 different methods:

1. pthread mutex

2. pthread spinlock
3. std atomic
4. Common atomic—an improvement to std atomic, where the scheduler thread signals all worker threads using a common atomic variable rather than an individual atomic variable per thread.

Table 4 and Table 5:

| | pthread mutex | pthread spinlock |
|---|---|---|
| variable | pthread_mutex v | pthread_spinlock v |
| lock() | pthread_mutex_lock(v) | pthread_spin_lock(v) |
| unlock() | pthread_mutex_unlock(v) | pthread_spin_unlock(v) |
| wait() | lock(); unlock() | lock(); unlock() |

**Table 4: Mutex and spinlock sync methods**

| | std atomic |
|---|---|
| variable | std::atomic<char> v; |
| lock() | v.store(1, memory_order_release) |
| unlock() | v.store(0, memory_order_release) |
| wait() | while ( v.load(memory_order_acquire) == 1 ) |

**Table 5: Atomic sync method**

Note the different implementations of wait(); in "std atomic" (Table 5) wait() is implemented using a loop, while in the other two implementations (Table 4) we used lock() followed by unlock(). ScaleSimulator may run up to N-1 worker threads on a server with N cores. Each worker thread does some work and waits on a barrier for all other workers to reach the common point before it can continue, and this is repeated cycle by cycle. The system (1) validates that all workers are working on the same iteration number and (2) counts how many loops each worker does in a second.

To measure the barrier efficiency, we conducted an experiment, where the simulator code has been manipulated to skip the actual work and transfer, leaving only the synchronization activity. We measured the barrier speed using Intel® Xeon® CPU E5-2660 v2 @ 2.20GHz, 20 cores / 40 threads [18]. The speed is measured in terms of the number of phases it can achieve in a second vs. the number of worker threads. The result is presented in Figure 9 below.

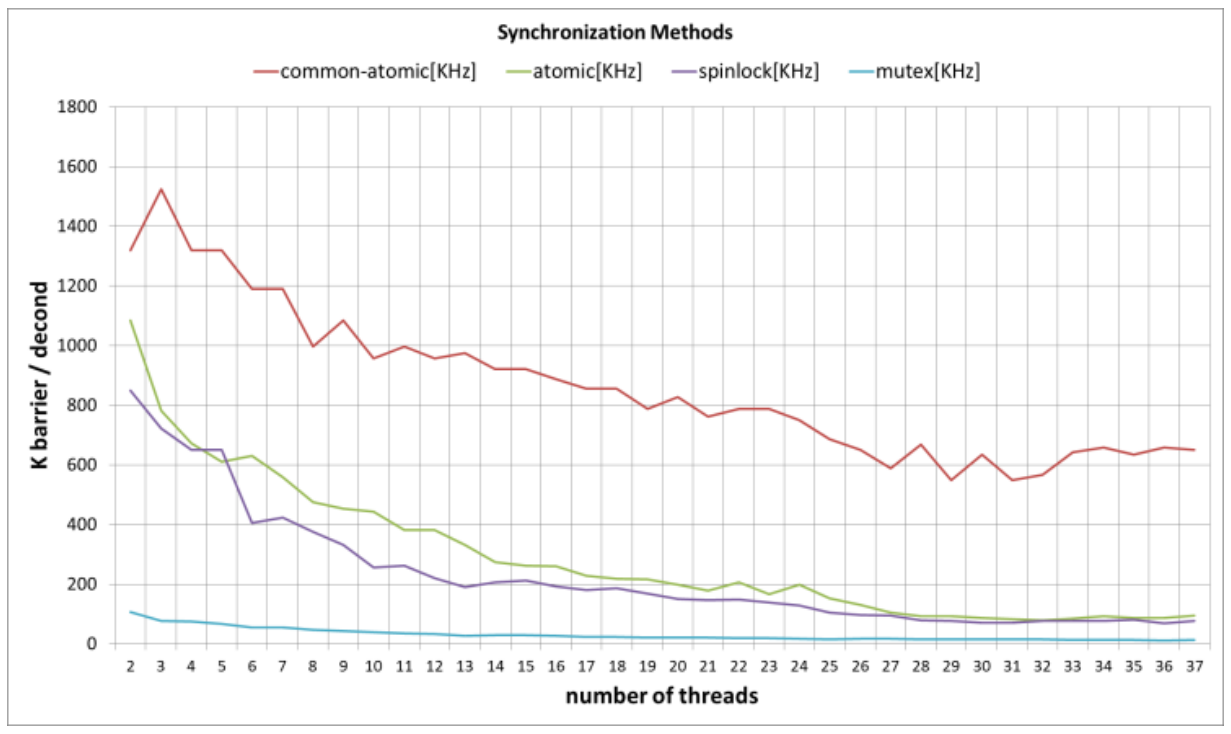

**Figure 9: Synchronization overhead**

As Figure 9 indicates, the common-atomic method outperforms all the others and yields very impressive numbers. We can also observe that the common-atomic method scales well and slows down only around 2x when moving from 2 worker threads to 37, while all the other methods degrade severely. The major differences between common-atomic and the other methods are as follows:

- In common atomic, a common sync-point is used by the scheduler to signal all worker threads.
- In the other methods, an individual sync-point is used by the scheduler to signal to each worker thread separately.

The level of parallelism in ScaleSimulator is limited by either the number of server cores or by the number of units in the model:

Maximum threads = min(server cores, model units). This high level of parallelism requires the software barrier to be very efficient and encourages the synthetic experiments presented above. In the next section we will examine the scalability of the solution when using larger numbers of cores.

We can compare our results to the scaling numbers as reported for the Hornet simulator in [9], where the authors report on the scaling of the simulator with respect to the number of threads and the frequency of synchronization points. Hornet is designed for multi-core simulation models only. The authors of Hornet identified that it is synchronization-limited and proposed to sacrifice some accuracy by synchronizing every 5 cycles or more. In contrast, ScaleSimulator is designed for a large variety of models and uncompromised cycle accuracy. The simulation results show that ScaleSimulator's parallel scaling with cycle accurate simulation can match and exceed Hornet's scaling with compromised multi-cycle synchronization. Other simulators, such as ZSIM [15], rely on speculative execution of parallel threads. As Figure 9 indicates, our simulator, when synchronized at every cycle, can still achieve an impressive speedup.

So far we have examined the implementation of different synchronization primitives using a moderate number of cores. To examine the impact of using a large number of cores, we took a larger machine that contains 8 sockets, with 24 cores per socket and 2 threads per core (a total of 384 threads), and measured the barrier speed again, using the common-atomic method.

Figure 10 shows a moderate degradation of the barrier speed from 8 to 256 threads.

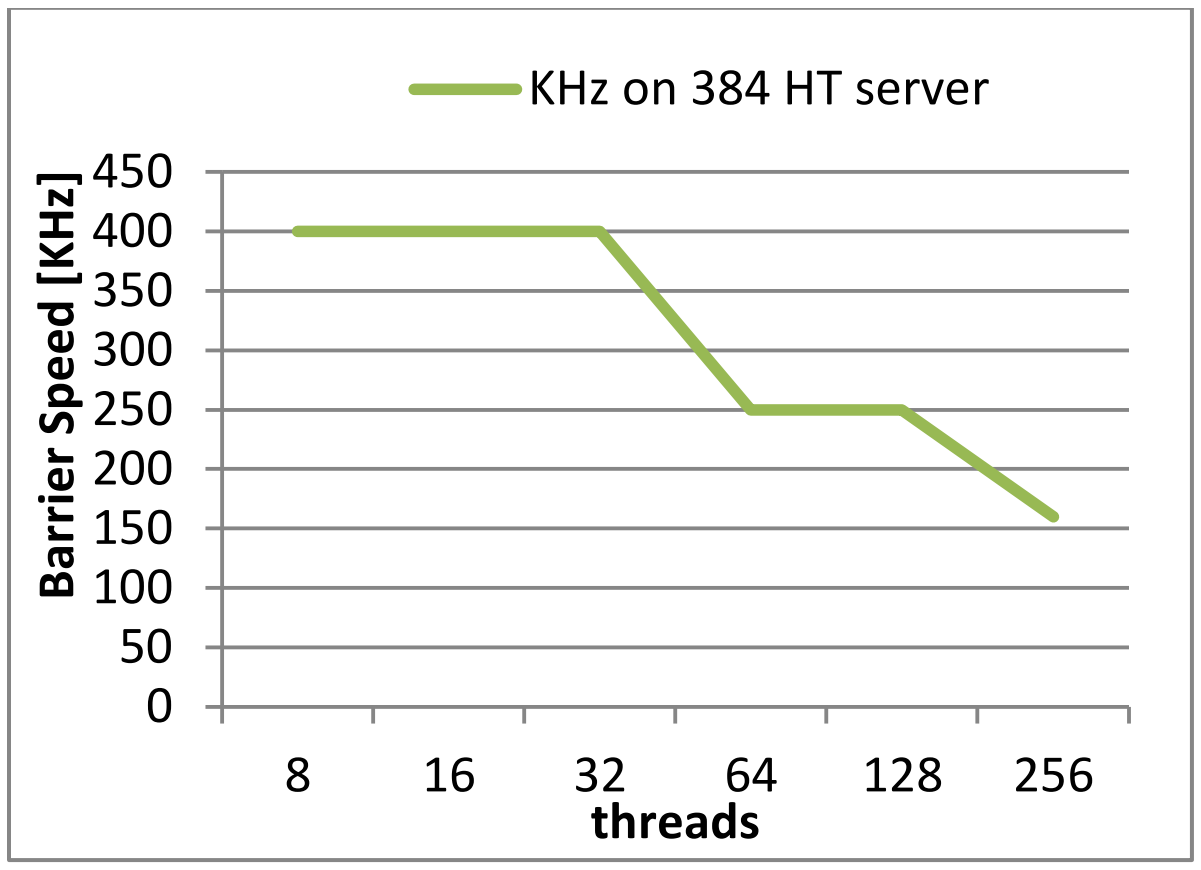

**Figure 10: Barrier speed on a 384 HT server**

As Figure 11 indicates, moving from 8 threads to 256 (32x) provides a 14x execution speedup.

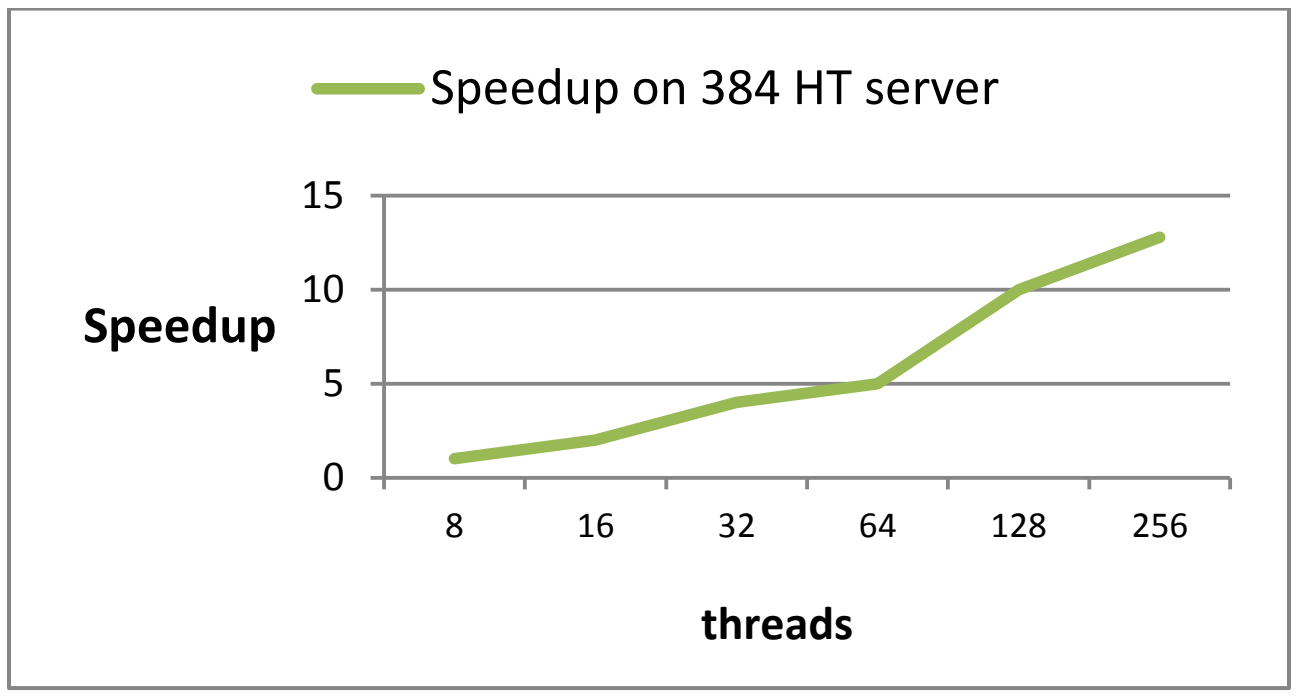

**Figure 11: Synchronization on a 384 HT server**

### 5.2 The use of OLTP based applications with a light CPU model

In the previous section, we used a synthetic workload to examine the impact of synchronization on overall system performance. In this section, our goal is to simulate a realistic workload that runs OLTP applications.

The simulated model consists of 32 light cores, light NoC, and each core has private L1 and L2 caches, and shared L3 with full coherency. The experiment assumes N worker threads, while the simulated cores are evenly distributed among them. For example, while using a single worker thread (serial simulation), the worker is in charge of simulating the entire 32 cores, whereas when using 16 worker threads, each working thread is in charge of simulating 2 cores.

Figure 12 compares the run of 5 different configurations. For each one we measure the overall execution time of the run (the blue bar), the relative time it takes to simulate a single cluster (2 cores) (red bar), and the overall time that the synchronization and overhead took (green bar). Please note that although the synchronization is done in parallel, the slowest worker thread dominates the simulation speed.

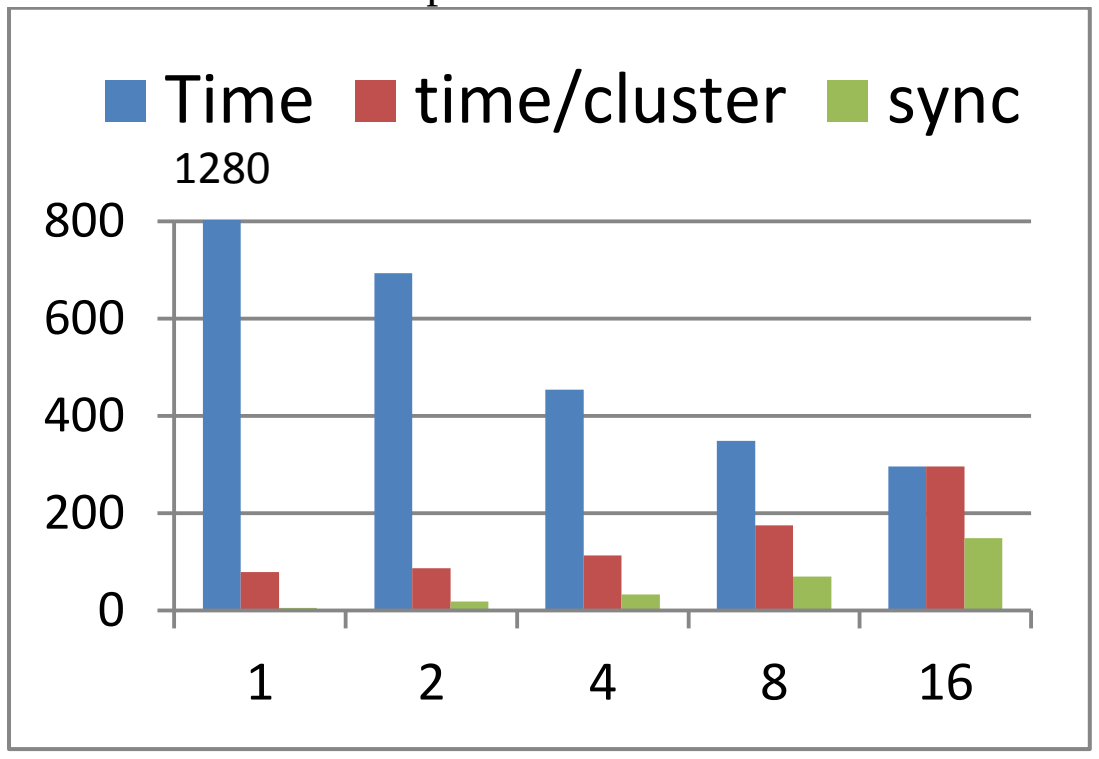

**Figure 12: OLTP light CPU simulation**

As Figure 12 shows, the system exhibits good scaling as the number of cores used to execute the simulation model increases. The experiment examines the simulation of 16 cores running OLTP, while using 1-16 threads (x-axis). As can be expected, when a single core is used to simulate the entire system, the overall execution time of the model is the longest, but we do not have any "serial related" overhead. On the other end of the scale, we are running each simulated core on a separate physical core. Here, the overall execution time is the best. As the overall simulation speed is very fast (greater than 100 KHz), the synchronization overhead of two barriers in a simulated cycle is not marginal.

To better understand the results, we performed another set of experiments. Here we measured how much the system spends in the "work phase" and how much the system spends in the transfer phase for each configuration.

As Figure 13 indicates, the time the simulator spends on the transfer phase remains almost the same in all configurations, but the time the simulator spends executing the work phase dramatically increases when the number of workers is large. At first glance, it may appear that the synchronization cost is the main reason for this increase, but Figure 9 shows that there must be another reason.

A closer look indicates that the dramatic increase in time is in fact due to the random distribution of the units. The cost of copying the messages between the different physical cores is actually paid during the work phase, when the receiver unit reads the transferred message and causes the server CPU cache coherency to perform a read-shared operation in order to move data to the receiver core. Hence this is the main limitation of the ScaleSimulator at that point.

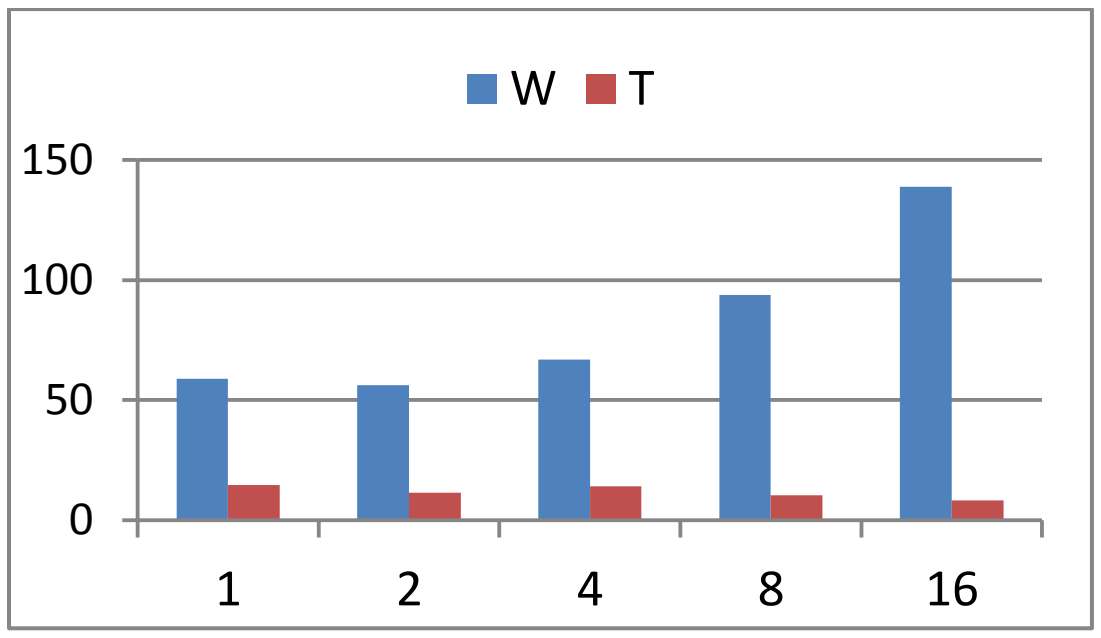

**Figure 13: Work vs transfer per worker**

### 5.3 Simulating out-of-order fully coherent systems

So far, when simulating a "real workload," our model was limited to simple cores only.

This section is devoted to the simulation of a cycle accurate model of a full CPU with 8 out-of-order cores and a cycle accurate NoC with full cache coherency running an "unmodified" OLTP benchmark. The results for these experiments are provided in Figure 14.

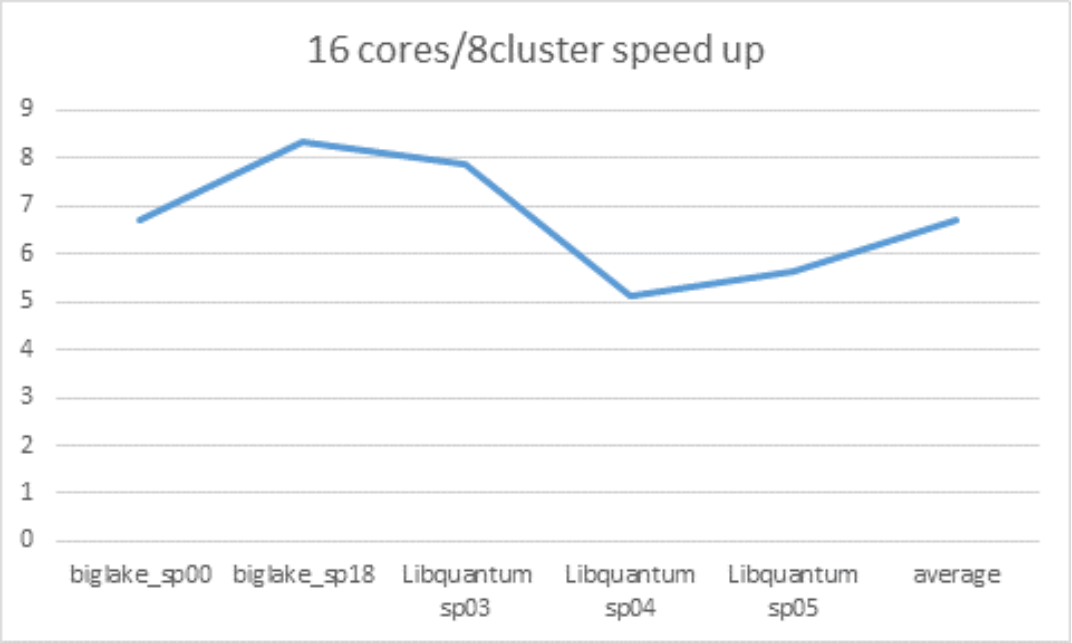

**Figure 14: Speedups of OOO based platform**

The results indicate that even when simulating complex core architecture and running OLPT based workloads, we can still get a sustainable speedup. In some cases the speedup slope is around 1, which means no parallelism penalty.

It should be noted that one major difference between the light and full CPU models is that full CPU runs 10-20 KHz per core, while light CPU runs 100s of KHz per core. The latter speed introduces two bounds:

1. The benefits of parallel simulation decrease due to data transfer between units running on different cores. If two units run on the same core, the data is transferred through the L1 cache of the server CPU. If the two units run on different cores, the data is transferred through L2 or L3. More parallel threads mean more data transfer through a higher-level cache, until the penalty exceeds the gain. At high simulation speed, the data rate is higher too and the bound is reached when running fewer threads.
2. As Figure 9 shows, the barrier synchronization speed is 100s of KHz, and if the overall simulation speed is beyond 100 KHz then the barrier synchronization overhead is no longer marginal.

### 5.4 Simulating a Data-Center model

This section describes a different type of simulation. Here, we use the ScaleSimulator to simulate cycle-accurate communication within a data center that contains 128,000 nodes connected through 5,500 switch devices, each of which has 128 ports. The switches are modeled to ascertain the level of accuracy, including their internal buffers, pipeline latency and the impact of the back pressure when resources are fully exhausted.

Unlike the previous cases where the workload was "generated" by running an application benchmark, this model uses a simple pseudo-random function to generate the source and the destination of 3,000,000 packets. The purpose of this experiment is to test the scalability of the system when using 1-24 physical cores to run the same simulator from start to end. We use this experiment to demonstrate the ability to scale an environment mainly governed by events.

Figure 15 presents the overall runtime of the simulator. We can observe that even for such a simulation environment the system can still scale in a reasonable manner.

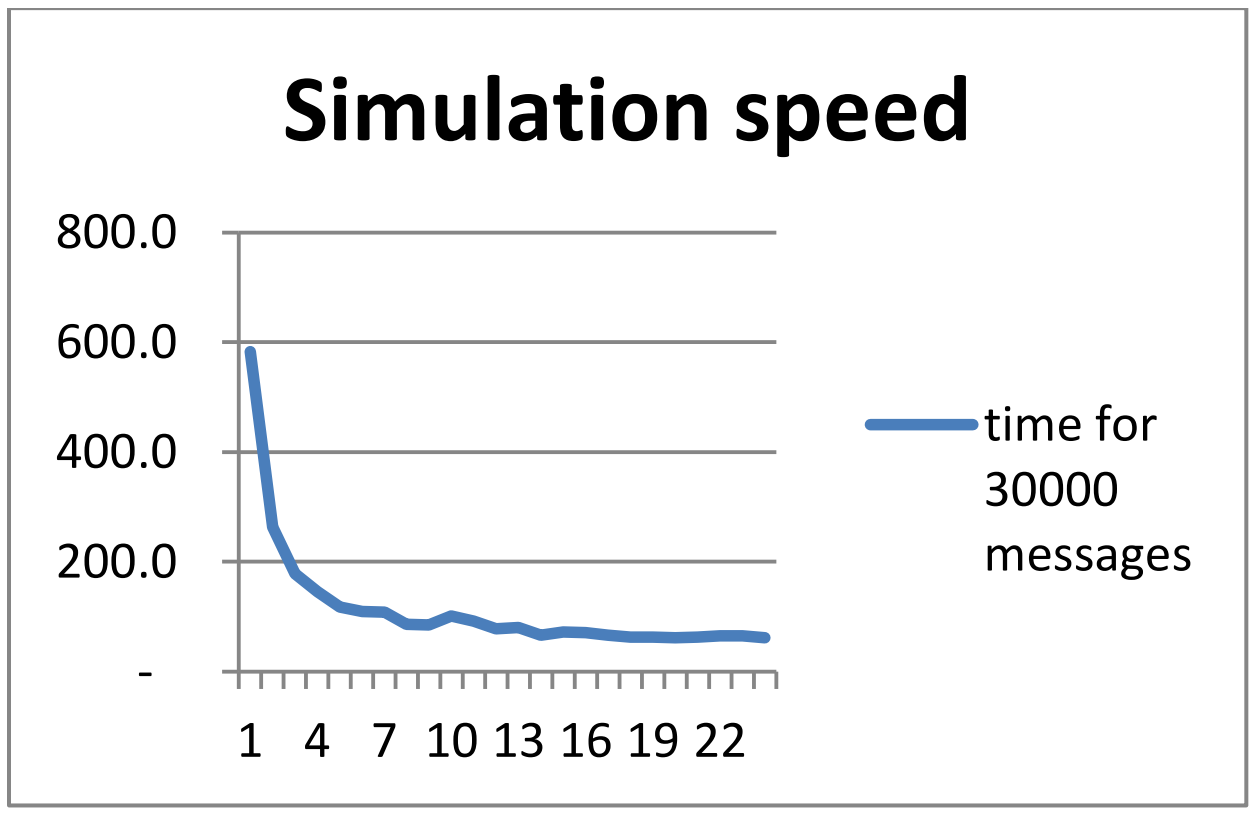

**Figure 15: Data center simulation**

Figure 16 presents the same simulation results but with an emphasis on the speedup time as compared to a sequential execution.

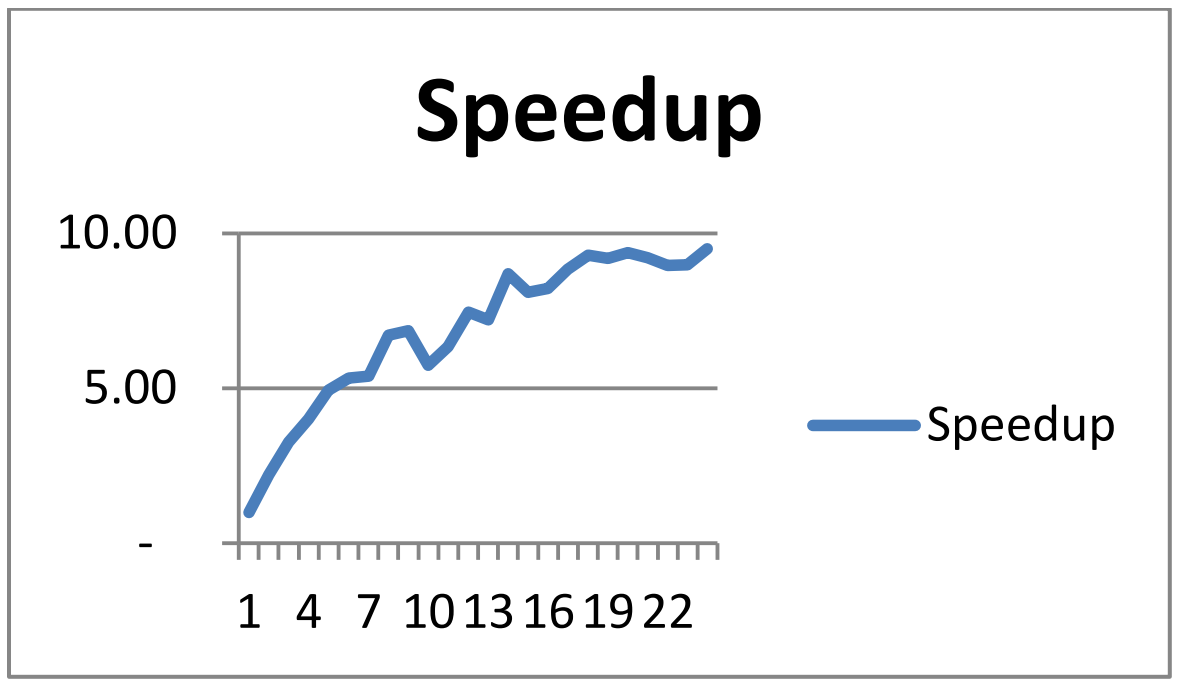

**Figure 16: Data center speedups**

Please note that a reasonable speedup of 6-10 times is achieved when we parallelize the simulations.

## 6 Conclusions and remarks

This paper presents a new approach for implementing cycle accurate simulations that can scale to a relatively large number of cores, thus allowing accurate simulation of future architectures that may have complex internal structures, using a large number of cores while still running meaningful benchmarks. The good scaling of the simulator is achieved by a new and unique technique of parallelization that uses a 2.5-phase execution. This allows

- Most of the operations to be performed in a thread-safe lockless data access.
- Efficient barrier implementation.
- A number of lock operations proportional to the number of simulating (physical) cores rather than to the size of the model or the number of data accesses.

We tested the new proposed simulation technique on realistic workloads, under various models with different levels of accuracy, and showed the benefits and the limitations of our new simulation environment. We believe that our new proposed ScaleSimulator is the first to accurately simulate cycle-by-cycle complex systems while still achieving an impressive scalability that allows it to run meaningful workloads.

Future work may focus on different optimizations to further improve the current implementation. One possible optimization is the distribution of units, which is currently random. We believe that a hierarchical ordering that will take advantage the locality and organize them accordingly could yield a significant improvement when simulating a large number of parallel units. The amount of data transferred between the server cores should also be optimized, as it leads to degraded performance due to cache coherency of the simulation server.